\documentstyle[11pt,twoside,colloqOHP2010,epsfig]{article}
\newcommand{\corot}{\emph{CoRoT}}

\begin{document}
\title{The CoRoT Exoplanet program : status \& results}
\author{M. Deleuil$^{1}$, C. Moutou$^1$, P. Bord\'e$^2$ \& the \corot\ exoplanet science team}
\affil{$^1$Laboratoire d'Astrophysique de Marseille, 38 rue F. Joliot-Curie, 13388 Marseille Cedex 13, France}
\affil{$^2$Institut d'Astrophysique Spatiale, B\^at. 121, 91405 Orsay, France} 

%
\begin{abstract}
The CoRoT satellite is the first instrument hunting for planets from space. We will review the status of the CoRoT/Exoplanet program. We will then present the CoRoT exoplanetary systems and how they widen the range of properties of the close-in population and contribute to our understanding of the properties of planets.
\end{abstract}
\section{Introduction}
\corot\footnote{The CoRoT space mission has been developed and is operated by CNES, with contributions from Austria, Belgium, Brazil, ESA, Germany and Spain.} has been designed with two distinct scientific objectives which both require continuous observations and ultra-high precision relative stellar photometry : one is the detection of extrasolar planets by the transit method; the second is the study of stellar interiors by asteroseismology  (Baglin et al. 2009). In addition, the long duration of the observation periods together with the exquisite photometric precision allow number of stellar physics studies. 
The instrument started the scientific observations on February 2nd 2007. The mission was initially scheduled for 3 years but extended for 3 additional years, that is till March 2013. 
Auvergne et al. (2008) give a complete description of the instrument in flight based on raw and calibrated data. In the present paper, we will just recall the main characteristics of the mission.
\section{CoRoT, brief summary}
\subsection{Photometry}
The typical exoplanet targets have a V-magnitude in the range $\simeq$11 up to 16. The photometry is performed on board for up to 6000 selected targets per CCD, 2 CCDs being dedicated to the exoplanet program. For a given target, the flux is thus integrated over a pre-designed photometric mask that encompassed the quite large PSF of the exoplanet channel, whose typical size is about 35 $\arcsec$ x 23 $\arcsec$. For stars brighter than magnitude 15, the pixels within the photometric mask are separated in 3 sub-sets according to the percentage of stellar flux. The co-addition of the pixels intensity over these 3 areas provides 3 different light curves, referred to as the red, the green and the blue light curves. 
\begin{table}
\begin{center}
\caption{\label{runs} \corot\ runs from 2007 to 2010.}
\begin{tabular}{lccrrrc}
\hline
Run & Date  & duration  & Nb of & Nb &  to  & Data  \\
       & start   &  [d]          & Targets & Cand. & FUp &  Status \\
\hline
$IRa01$ & 02/2007 & 45 &  9 921 & 254 & 40 & Public \\
$SRc01$ & 04/2007 & 26 & 7 015 & 261 & 62 & Public \\
$LRc01$ & 05/2007 & 152 &  11 448 & 229 & 29 & Public \\
$LRa01$ & 10/2007 & 150 &  11 448 & 304 & 79 & Public\\
$SRa01$ & 03/2008 & 25 &  8 189 & 163 & 29 & Public\\
$LRc02$ & 04/2008 & 150 &  11 448 & 286 & 56 & Public\\
$SRc02$ & 09/2008 & 21 &  11 448 & 336 & 45 & Public\\
$SRa02$ & 10/2008 & 32 &   10 305 & 217 & 32 & Public\\
$LRa02$ & 11/2008 & 115 &   11 448 & 362 & 36 & Public\\
$LRc03$ &04/2009 & 89 &  5 724 & 244 & 61 & Public\\
$LRc04$ & 07/2009 & 83 &  5 724 & 173 & 51 & CoIs \\
$LRa03$ & 10/2009  & 148 &  5 329 & 124 & 22 & CoIs \\
$SRa03$ & 03/2010 & 24 & 4 169 & 100 & 15 & on alarm \\
$LRc05$ & 04/2010 & 84 & 5 724 & 89  & 22 & on alarm\\
$LRc06$ & 07/2010 & 77 & 5 724 & 100 & 21 & on alarm\\
\hline
\end{tabular}
\end{center}
\begin{footnotesize}
Column 5: number of transit candidates detected; Column 6: number of candidates assessed worth for follow-up observations. Following the DPU 1 break down in March 2009, the number of light curves is divided per 2 in the last fields.  
\end{footnotesize}
\end{table}
These light curves have however no direct correspondence to any standard photometric system. For fainter stars, only the standard single-band photometry (white) is performed.  The integration time on board is 32 $sec$ but the flux of 16 read-outs is coadded on board over an 8.5 $min$ time span before being downloaded. The elementary 32 $sec$ integration time is however preserved for selected targets, whose list is regularly updated during \corot\ observations thanks to the {\sl Alarm mode} (Surace et al. 2008). In march 2009, one of the two DPUs which pilot the on-board photometric treatment had a break down and could not be switched in operation again. As a consequence, the field-of-view for exoplanet search is yet half-reduced to (1.35$^o$)$^2$. 
\begin{figure}
\begin{center}
\includegraphics[scale=.18]{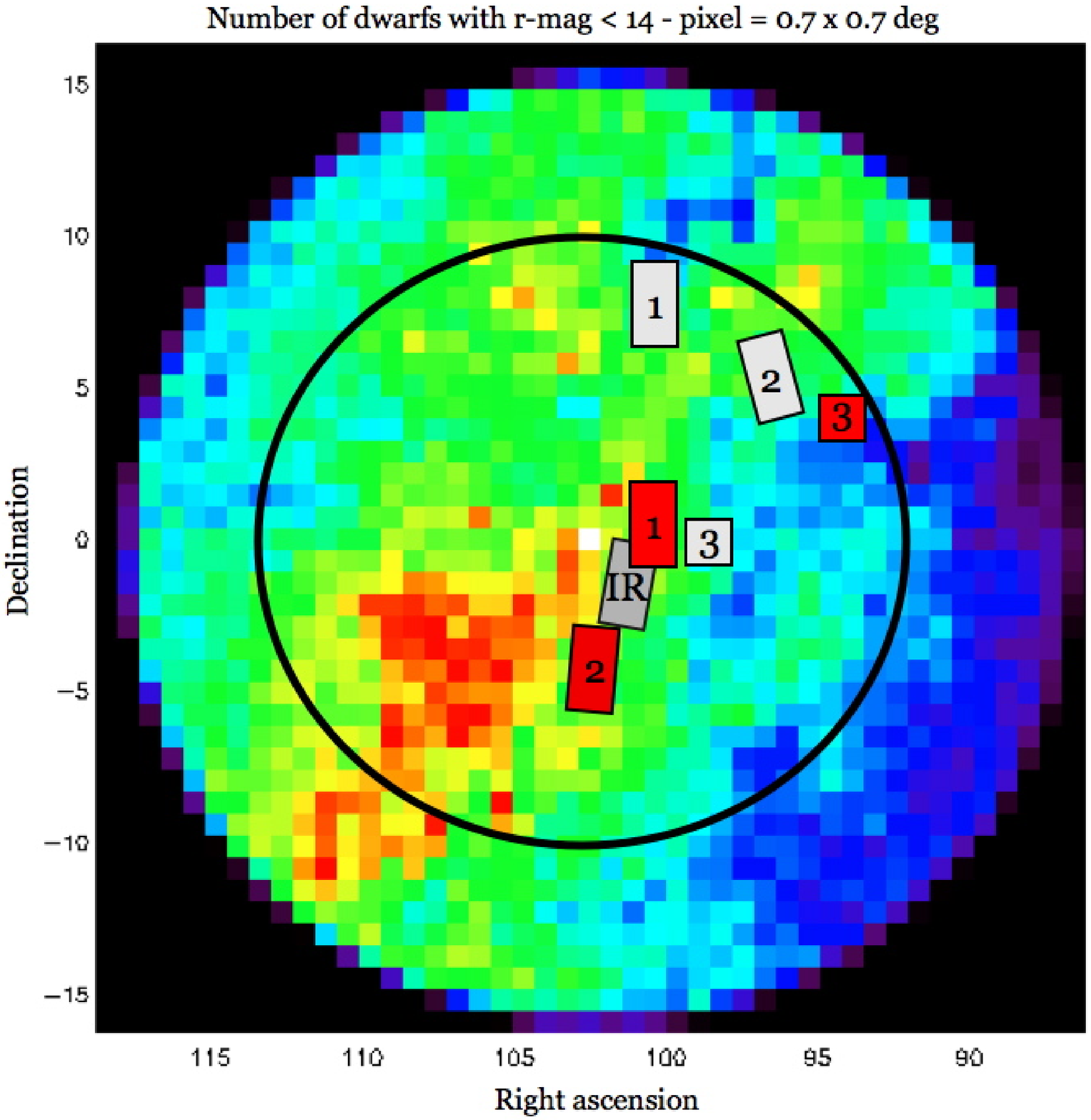}\includegraphics[scale=.18]{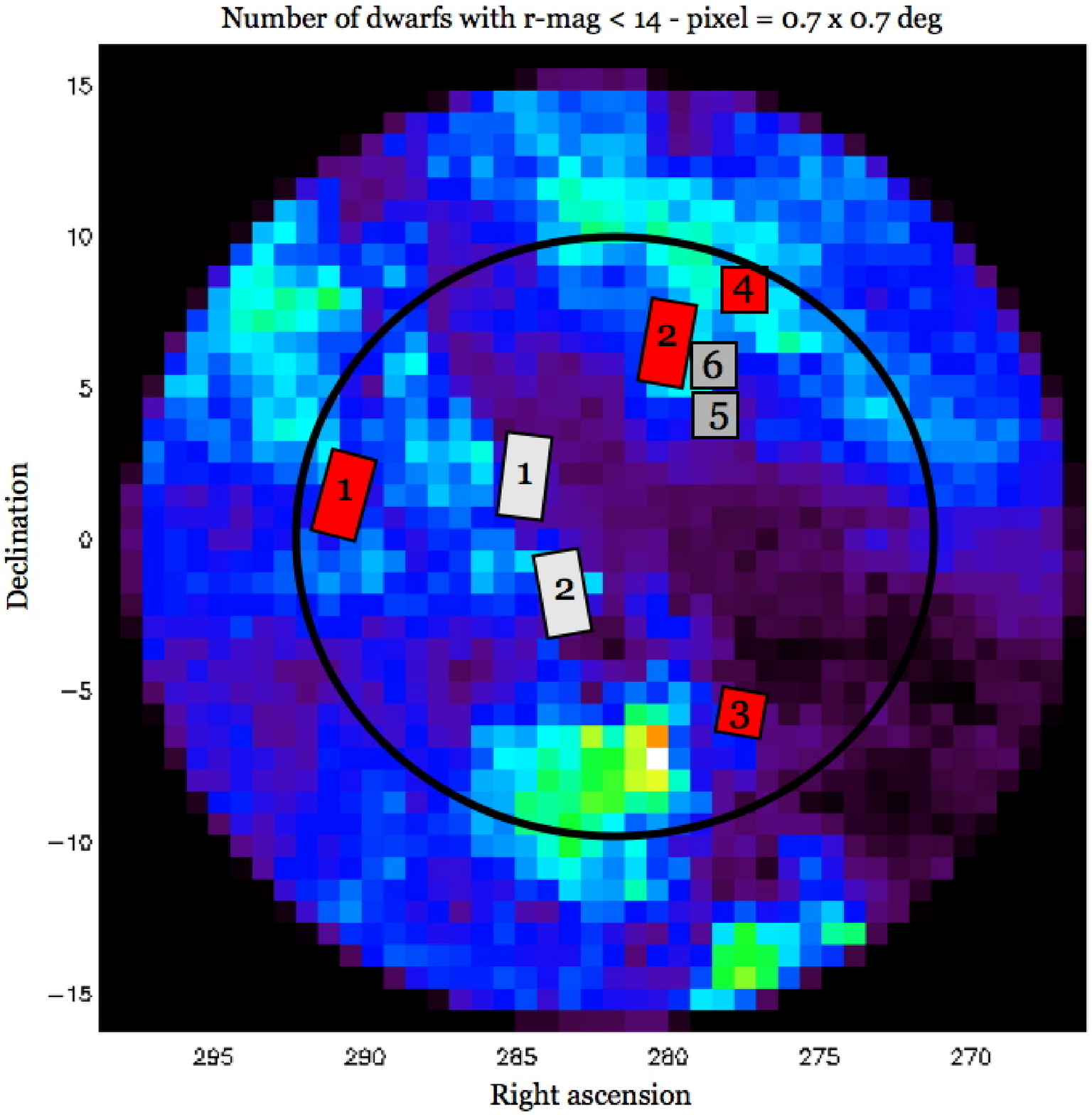}
\caption{The two continuous pointing zones of \corot. The small rectangular and square shapes show the exoplanet fields that have been observed up to summer 2010. Note the extreme reddening variations towards the center fields.}
\end{center}
\end{figure}
\subsection{Observation strategy}
The instrument has 2 continuous viewing zones which are two circles of $\simeq$10$^o$ of radius, centered in the galactic plane and separated by 180$^o$ in right ascension.This ensures a level of background light due to the Earth reflected light as low as possible. In addition, in order to avoid to be blinded by the Sun, twice a year the instrument is rotated and is pointed in the opposite direction. The strategy of observation is flexible but consists mainly in one long run (LR) lasting for about 140 days and one shorter run  (SR) whose duration is between 20 and 30 days per half-year. The fields observed in the circle centered at  6$^h$ 50$^m$ in right ascension are referred as "anti-center fields", and those in the circle at 18$^h$ 50$^m$, as "center fields". Table~\ref{runs} gives a summary of the stellar fields that have been observed since the beginning of the operations. With only one CCD left in operation, the strategy of observation has been adapted and some of the long runs have been replaced by two runs of a shorter duration. \\
Figure 1 shows the location of the various \corot\ exoplanet fields inside the two \corot\ continuous viewing zones overplotted on a dwarf density map. The two regions have different stellar content in terms of dwarf and giant but also in their distribution over the various spectral type (Gazzano et al. 2010). This allows to probe regions in the galactic plane with different stellar properties in terms of metallicity but also in terms of age. Once the statistics of planets found by \corot\ will be much more significant, this should bring some insights into the physical conditions that prevail for planet formation mechanisms.

\section{Detection capacity}
\begin{figure}
\begin{center}
\epsfig{file=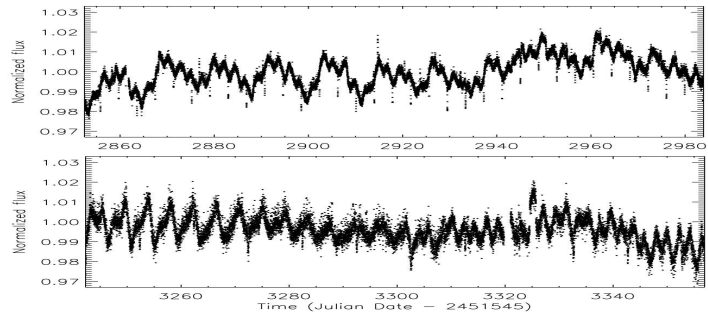,height=120pt,width=1.0\linewidth}
\caption{Two examples of \corot\ light curves .}
\end{center}
\end{figure}
In summer 2010, a total of 129 326 light curves was collected (Table~\ref{runs}). These light curves show a fascinating diversity with various kind of stellar behavior: quiet stars display nearly constant a flux while others exhibit a strong variability over a time scale of a few minutes (Fig. 2). All these light curves are however analyzed for transit searches. Transits are detected in about 100 up to 300 light curves per run, that is in about 3000 light curves for the whole mission to date. They present various depth, shape and duration. These detected transits are however predominantly transiting stellar systems. About 80\% to 90\% of these stellar systems are pinpointed thanks to the long duration of the light curves, through the identification of the secondary transits or light curve modulation.  The remaining candidates are screened out by follow-up observations, using various techniques from photometric observations to radial velocity measurements. These follow-up observations play a major role in the \corot\ science and require a huge observational effort. They allow to assess the nature of the detected transiting body, stellar or planetary and, in the later case, to measure its mass. Bouchy et al. (2009) give a complete overview of the strategy for radial velocity observations and Deeg et al. (2009) present the ground-based photometric follow-up ones.
\begin{figure}[t]
\begin{center}
\epsfig{file=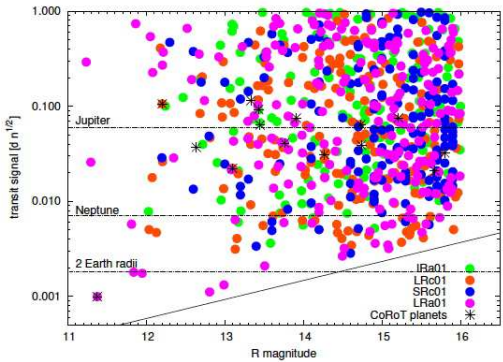,width=0.6\linewidth}
\caption{Transit signal for all candidates detected in the \corot/exoplanet fields observed during the first year (adapted from Cabrera et al.(2009)), with the expected signals for a Jupiter, Neptune and Earth-size like planets respectively indicated.}
\end{center}
\end{figure}
Whatever their nature, stellar or planetary, one can use the detected transits to assign the performance of the transit detection from the \corot\ light curves. Figure 3 shows the transit signal defined here as the product of the relative depth of the transit and the square root of the number of points in the transit, as a function of the target magnitude. The colored circles are the candidates identified in the various runs according to the color codes in the bottom right and the black stars the planets. It shows that Neptune size planets can be detected whatever the magnitude of the star is, whereas super-Earth size planets, such as CoRoT-7b can be discovered in the light curves of stars brighter than R $\simeq$ 14 only. 
\section{The CoRoT planets }
Table~\ref{Pln} gives the main characteristics of the \corot\ planets published by the end of summer 2010. These planets correspond to the first discoveries in the first two years of data. A large fraction of the candidates are indeed still in the screening process by follow-up observations. Their properties, compared to the other published exoplanets are illustrated in Fig  4. The mass-radius diagram shows that most of these planets  are in the domain of gaseous giants but the long duration of the light curves and the high duty cycle greater than 90\% of space-based observations allow to explore the transiting systems over an extended orbital domain. The \corot\ discoveries indeed account for half of the transiting systems with an orbital period greater than 8 days.
\begin{table}
\begin{center}
\caption{\corot\ discoveries main characteristics.}
\label{Pln}
 {\scriptsize
\begin{tabular}{cccccccccl}
\hline
Ref & Run &  Radius     & Mass         & Density & Period & Nb & Sp. & Star & Comment   \\
   &         & R$_{Jup}$ & M$_{Jup}$& $g cm^{-3}$ & days & & Type & [Fe/H] &  \\
\hline
1 & IRa01 & 1.49 & 1.03 & 0.34  &  1.509 & 34 & G0V & -0.20  & bloated HJ \\
2 & LRc01 & 1.46 & 3.31 & 1.31  &  1.743 & 78 & G7V &  0.00   & bloated HJ  \\
3 & LRc01 & 1.01  & 21.6  & 26.4   &  4.26 & 34 & F3V &  -0.02   & BD or planet   \\
4 & IRa01 &  1.19  & 0.72 & 0.52  &  9.20 & 7 & F8-9V & +0.05   & synchronized  \\
5 & LRa01 &  1.39  & 0.46 & 0.22  &  4.04 &  27 & F9V &  -0.25 & metal poor HS   \\
6 & LRc02 &  1.16  & 2.96 & 2.32  &  8.89 & 15  & F9V &  -0.20  & metal poor HS  \\
7 & LRa01 & 0.14   & 0.007-0.02  & 3.4-9.6  & 0.85   &  153     &  G9V & 0.12  & Super-Earth  \\
8 & LRc01 &  0.57  & 0.22 & 1.6   &   6.21    & 33  & K1V &  0.30  &  Icy giant    \\
9 & LRc02 &  1.05  & 0.84 & 0.90   & 95.27  &  2     &  G3V & -0.01 &  temperate   \\
10 & LRc01 &  0.97  & 2.75 & 3.70 & 13.24  &  10    & K1V &  0.26   & eccentric   \\
11 & LRc02 &  1.43  & 2.33 & 0.99 & 2.99  &   49   & F6V &  -0.03   & $v sini$ = 40 km/s   \\
12 & LRa01 &  1.44  & 0.92 &  0.31 & 2.83  &   47  & G4V &  0.16  & bloated low density HJ   \\
13 & LRa02 &  0.9  & 1.3 & 2.34 & 4.04  &  31   & G0V &   0.01  & dense HJ  \\
14 & LRa02 & 1.09  & 7.6 & 7.3  & 1.15   &   89 & F9V  &  0.05   & extreme density  \\
15 & SRa02 &  1.12  & 63.3 & 59 & 3.06  &  10    & F7V &  0.10  &  BD \\
\hline
\end{tabular}
}
\end{center}
\begin{footnotesize}
The planets are ordered by increasing \corot\ number. HJ stands for Hot Jupiter and HS for host star. Column 7 gives the number of transits in the \corot\ light curve.(1) Barge et al. (2008); (2) Alonso et al (2008); (3) Deleuil et al (2008); (4) Aigrain et al (2008); (5) Rauer et al (2008); (6) Fridlund et al. (2010); (7) L\'eger et al. (2009); (8) Bord\'e et al. (2010); (9) Deeg et al. (2010); (10) Bonomo et al. (2010);  (11) Gandolfi et al.(2010); (12) Gillon et al. (2010); (13) Cabrera et al. (2010); (14) Tingley et al. (2010); (15) Bouchy et al. (2010).
\end{footnotesize}
\end{table}
\subsection{Extending the orbital domain of transiting planets}
The first discoveries, CoRoT-4b and 6b are apart from the pile up of the  period histogram below 5 days (Fig. 4). In the case of CoRoT-4b, the continuous photometry allow to measure the rotation period of the star and to establish that it is a synchronized system (Aigrain et al. 2008; Lanza et al. 2009). Among the \corot\ planets, it remains the only system for which the synchronization could be clearly detected. Being not synchronized despite a similar orbital period, makes CoRoT-6b a good case to probe possible star-planet magnetic interactions. Lanza et al. (2010) used the light curve to map the longitudinal distribution of the photospheric active regions and trace their evolution. They reported some statistical evidences of active regions lagging the sub-planetary point during some temporal intervals. While the detection is still marginal, it is however a good illustration of the interest of uninterrupted sequence of photometric observations to probe the star - planet interaction and the evolution of planetary systems. 
\begin{figure}[h]
\begin{center}
\begin{tabular}{ll}
\epsfig{file=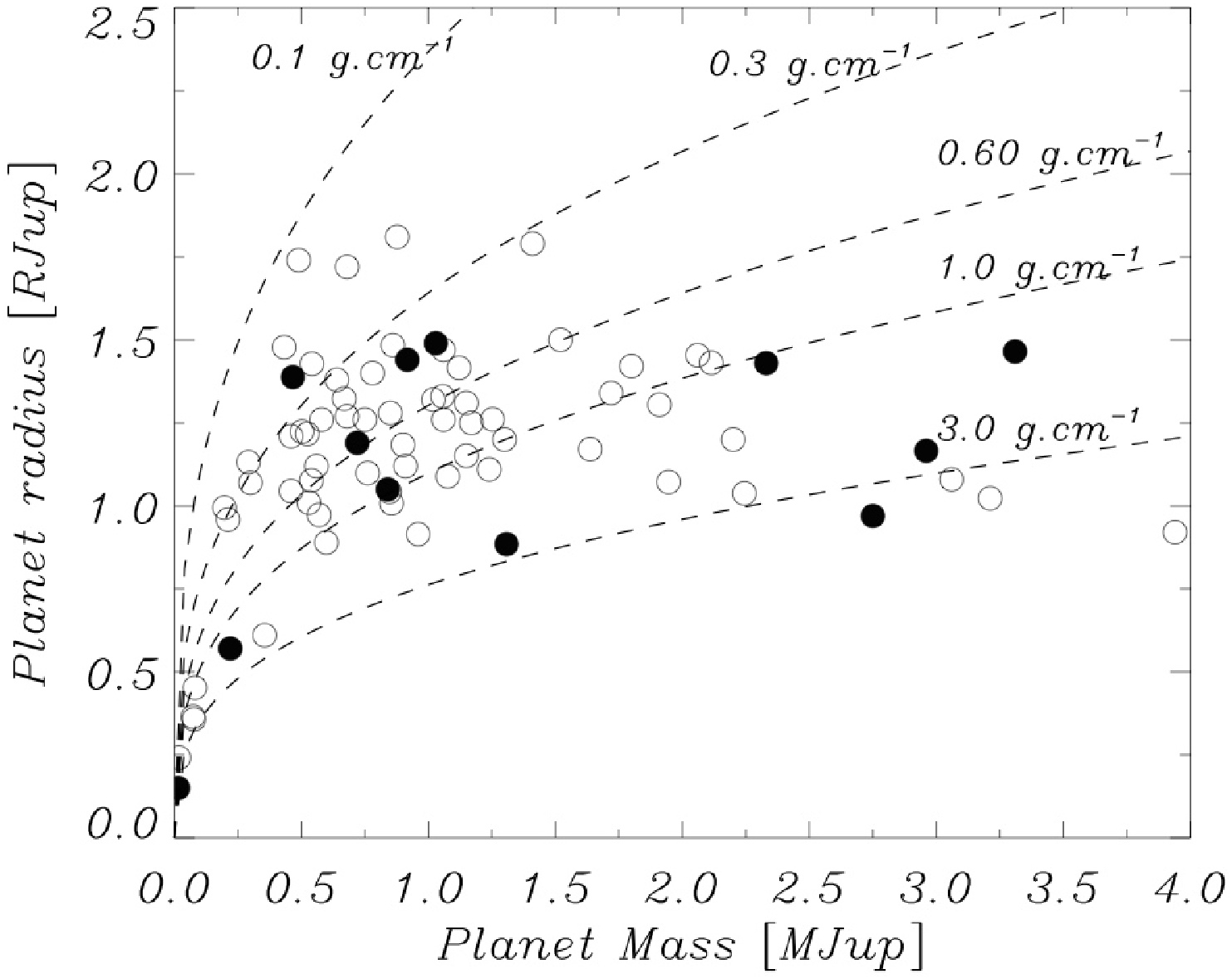,width=0.43\linewidth} 
\epsfig{file=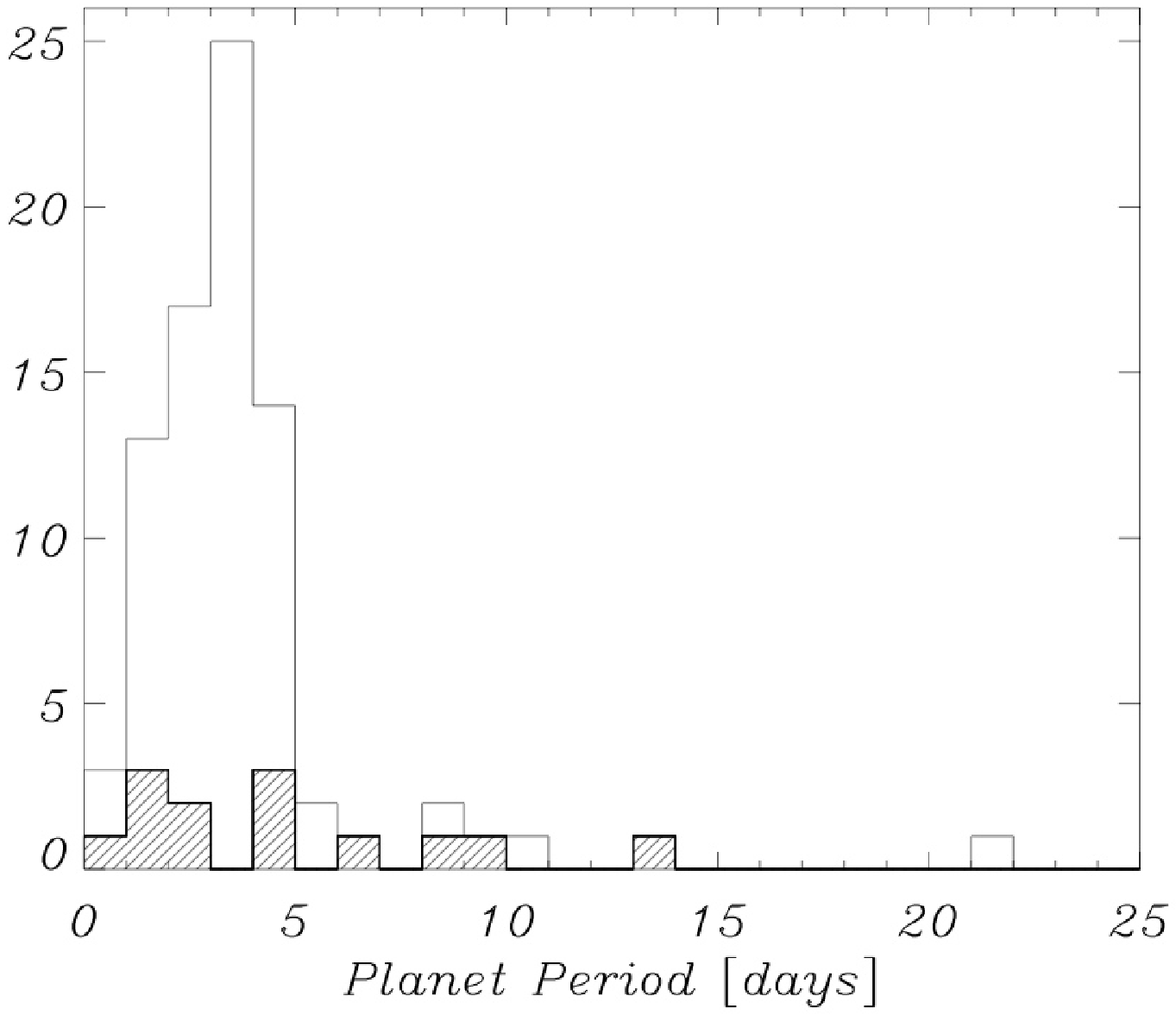,width=0.40\linewidth}
\end{tabular}
\caption{Mass - radius diagram for the transiting planets. Filled circles and the hashed histogram are the \corot\ planets.}
\end{center}
\end{figure}
With an orbital period of 13 days, CoRoT-10b belongs to the class of the few transiting exoplanets with highly eccentric orbit (Bonomo et al. 2010). The large variation in the orbital distance along the planet's year, results in a tenfold increase in the amount of stellar radiation received by the planet. Its eccentricity could have been produced by Kozai oscillations with a distant companion of stellar nature or by planet - planet scattering. No hint of non-transiting companion has been reported so far but clearly this system would deserve radial velocity monitoring in order to put further constrain on the origin of the planet's eccentricity. 

CoRoT-9b is the first long orbital period transiting planet discovered by a transit survey (Deeg et al. 2010). The planet orbits its solar like G3-type star in 95 days. Unlike the two other long-period transiting planets, HD 17156b (Barbieri et al. 2007) and HD 80606b (Moutou et al. 2009), this Jupiter's size planet has a low eccentricity. It makes it a perfect representative of the largest known population of planets (Fig. 5). CoRoT-9b is a very worthwhile object for further deeper studies to probe the properties of this exoplanet population.  
\subsection{Populating the brown dwarf desert }
\label{BD}
Exploring the exoplanet population at short orbital periods \corot\ has unexpectedly started to populate the so called brown-dwarf desert. 
While, with a mass of 21.6 M$_{Jup}$ for a JupiterÕs size, the exact nature of CoRoT-3b could be questioned as it lies in the overlapping region between the planet population and the low mass star regime (Deleuil et al. 2008), CoRoT-15b appears as a bona fide brown dwarf (Bouchy et al. 2010). Just as the most massive planets discovered so far, such as HAT-P-2b or CoRoT-14b (Tingley et al. 2010),  these two massive companions orbit an F-type star with a marked $v sini$ value. The significant rotational line broadening of the host star could be the cause of the lack of previous discovery of such massive companions from radial velocity surveys. It would have affected the search strategy of radial velocity surveys, either by removing these objects after one measurement or by setting them aside with a very low observation ranking priority. This would have resulted in a selection bias towards slow rotators that transit surveys have started to remove, allowing to enlarge the space of parameters for host stars. It reinforces the trend that the close-orbiting companions to host stars with mass above 1.1 M$_{\sun}$ could be more massive than companions to lower mass hosts. Probing the physical properties of such objects
could bring the missing clues to the understanding of the link between the population of planets and low-mass stars.
\subsection{Opening the small-size planet domain }
\label{HJup}
\begin{figure}
\begin{center}
\epsfig{file=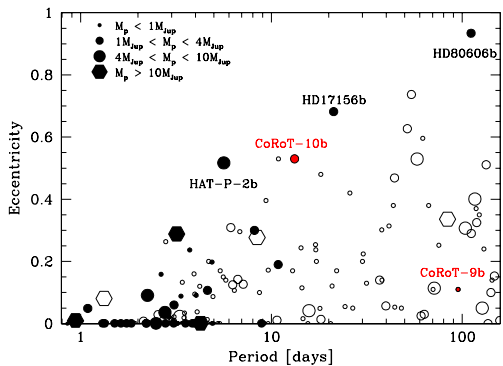,width=0.55\linewidth}
\caption{Eccentricity-period diagram for the known extrasolar planets (adapted from Bonomo et al. (2009)). Black filled symbols are transiting planets while open one are those with radial velocity measurements only. The size of the symbol indicates the mass range. }
\end{center}
\end{figure}
At the other extremity of the mass function, the recently discovered CoRoT-8b has a size in between Saturn's and Neptune's one (Bord\'e et al.  2010) but a density comparable to that of Neptune. It suggests a massive core with a mass in the range 47 up to 64 M$_{\oplus}$ and a much smaller H - He envelope. Like HD 149026b, the parent star is metal rich, confirming the suspected trend for the formation of giant planets with high content of heavy elements for metal rich  stars. In the mass - period diagram this planet lies in between the hot-Jupiter and the Super-Earth population, in a region where formation models predict a lack of planets. It might belong to the distribution tail of gaseous giants but clearly, with the intensive exoplanet searches, it would be interesting to see if this region of the diagram remains so sparsely populated in agreement with the bimodal distribution. 

Even smaller is CoRoT-7b, first planet discovered in the domain of Super-Earth size planets. The planet orbits in 0.8 days a K0 main sequence star which is one of the brightest star in the first long run in the anti-center direction. The transit is 0.35 $mmag$ deep but is clearly detected in the Fourier transform of the light curve with all its harmonics. The star's light curve is modulated by spots at the surface of the star dragged by the  star's rotation. This stellar activity produces a stellar jitter that competed the radial velocity measurements. More than 100 hundred HARPS spectra were collected over an one-year period. Different methods have been used to analyze the radial velocity signal and understand the stellar activity (Queloz et al. 2008; Hatzes et al. 2010; Boisse et al. 2010; Pont et al. 2010). Whatever the study, they confirmed the planetary nature of CoRoT-7b with a mass that appears consistent with a rocky planet but a poor precision, larger than 20\%. Not only the exact mass is suject to controversy but also the exact number of planets in the system. There are indeed some evidences in the radial velocity signal for a second non-transiting planet (Queloz et al. 2008) and even a third (Hatzes et al. 2010). Clearly to depict this system and accurately measure the mass of CoRoT-7b, additional velocity measurements are required with an adapted observation strategy so that the stellar activity imprint could be properly modeled. 
 
\section{Conclusion}
For more than 3 years \corot\ has been collecting high precision photometry light curves. The first discoveries from the analysis of the first two years of data proved the instrument capability to fully explore the domain of extra-solar planets at short orbital periods and detect planets over a wide range of size and properties from the first Super-Earth, CoRoT-7b, to CoRoT-9b the first temperate hot Jupiter. Continuous observations over long time spans also demonstrate their potential to probe different aspects of the planetary systems with, for example, the search for planet - host stars interactions.  With a few tenths of planet candidates in the follow-up process and new light curves to be analyzed and acquired in the two forthcoming years, new exciting results can be expected. 
   
\end{document}